\documentclass{moriond}

\def\be{\begin{equation}}
\def\ee{\end{equation}}
\def\bea{\begin{eqnarray}}
\def\eea{\end{eqnarray}}

\def\kpn{K^+\rightarrow\pi^+\nu\bar\nu}
\def\klpn{K_{L}\rightarrow\pi^0\nu\bar\nu}
\newcommand{\tev}{\, {\rm TeV}}

\newcommand{\Photo}{\includegraphics[height=35mm]{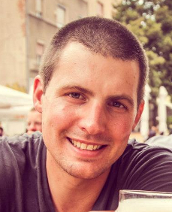}}

\begin{document}

\begin{flushright}
    {FLAVOUR(267104)-ERC-98}
\end{flushright}

\vspace*{4cm}
\title{$K \to \pi\nu\bar\nu$ in the Standard Model and Beyond}

\author{R.~Knegjens}

\address{\
TUM Institute for Advanced Study, Lichtenbergstr. 2a, \\D-85747 Garching, Germany}

\maketitle\abstracts{\
The precision expected for the rare $K\to\pi\nu\bar\nu$ decays by the NA62 and KOTO experiments in the coming decade will rival their current SM predictions.
In preparation for this upcoming opportunity, we review the SM predictions and discuss the sensitivity of these decays to models beyond the Standard Model,
considering in particular simplified $Z$ and $Z'$ models as benchmarks.
In the latter case we also discuss how these decays could ultimately probe distance scales as small as zeptometers i.e.\ peek into the Zeptouniverse.
}

\section{Introduction}

Since the turn on of the LHC half a decade ago, the high energy physics community has yet to discover a conclusive signal of New Physics (NP), its coveted goal.
However, it has narrowed down the search; in particular placing lower mass bounds on several hypothesized NP particles via direct searches.  
And much progress has also been made on indirect searches, which may be our last hope in the LHC-era should the NP scale prove to be out of reach of direct searches.
By indirect searches we refer in particular to flavour changing neutral currents (FCNC) processes,
which are necessarily loop suppressed in the Standard Model (SM) due to the GIM mechanism, and further so by the almost diagonal CKM matrix structure.
In contrast, FCNCs in models of NP need not be suppressed at all.
A prominent example is meson mixing, which is driven by $\Delta F=2$ FCNC processes that can probe NP scales up to thousands of TeV~\cite{Charles:2013aka,Bona:2007vi} if the NP is unsuppressed, or, equivalently, down to distances smaller than a zeptometer.
The catch is that were NP detected through such channels, many details would remain hidden.

This is where rare $\Delta F=1$ FCNC decays enter, which have the advantage that their operator structure, for example whether they couple left (LH) or right handedly (RH) to quarks or leptons, is exposed by the spin structure of the final state.
This would reveal much about the nature of the NP, and it is thereby worth asking what scales could ultimately be reached by such processes.
The two famous examples to be discussed in this talk are the rare decays $K^+\to\pi^+ \nu\bar\nu$ and $K_{\rm L}\to\pi^0 \nu\bar\nu$.
These decays are driven by electroweak (EW) loops, in particular $Z$-penguins with internal top quarks.
The two decays differ only by their spectator quarks, though due to its CP-even final state the $K_{\rm L}\to\pi^0 \nu\bar\nu$ decay is almost completely CP violating, in contrast to $K^+\to\pi^+ \nu\bar\nu$, which also has a CP conserving component.
Strikingly, the CKM structure corresponding to the leading $s \to d$ transitions in these decays is two orders of magnitude smaller than for $b\to \{d,s\}$ transitions, making these decays exceptionally suppressed in the SM.
Furthermore, the long-distance physics described by the hadronic matrix elements of these decays, which is typically a troublesome source of uncertainty for meson decays, can be accurately related using chiral perturbation theory to those of charged semileptonic decays~\cite{Mescia:2007kn}.
Thus these decays are also exceptionally theoretically clean, and thereby ideal probes of NP\@.

Experimental progress to date has been modest, with an imprecise branching ratio measurement for $K^+\to\pi^+ \nu\bar\nu$~\cite{Artamonov:2008qb}, and an upper-bound on $K_{\rm L}\to\pi^0 \nu\bar\nu$~\cite{Ahn:2009gb}.
It is therefore exciting that within the next 10 years the NA62 experiment at CERN hopes to measure the former mode with a precision of 10\% relative to the SM~\cite{Rinella:2014wfa,Romano:2014xda}, and the KOTO experiment hopes to observe the latter mode~\cite{Shiomi:2014sfa}.
Unfortunately, two experiments planned at Fermilab to measure both to a 5\% precision, ORKA~\cite{E.T.WorcesterfortheORKA:2013cya} and Project X~\cite{Kronfeld:2013uoa}, do not look set to continue.

Nonetheless, a lot is possible with the planned precision.
In Section~\ref{sec:SM} we will discuss the status and perspectives of these two decays in the SM. 
In Section~\ref{sec:BSM} we will discuss how their interplay can discriminate between various models of NP using simplified $Z$ and $Z'$ models as a basis.
Furthermore we will briefly discuss what NP scales could ultimately be reached in general for these decays.

\section{$K\to \pi\nu\bar\nu$ in the Standard Model}\label{sec:SM}

\begin{figure}[t]
\begin{minipage}{0.5\linewidth}
\centerline{\includegraphics[width=0.7\linewidth]{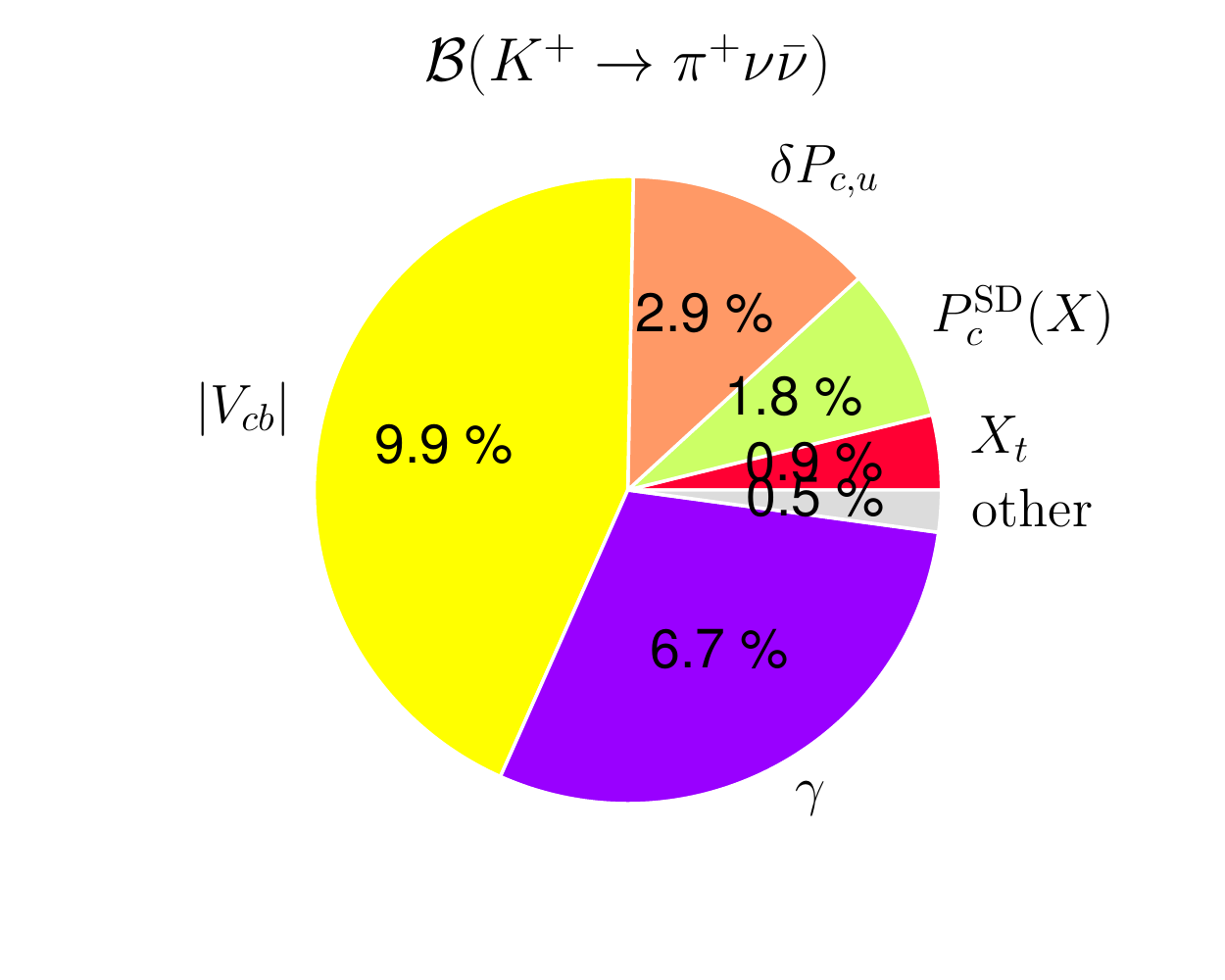}}
\end{minipage}
\hfill
\begin{minipage}{0.5\linewidth}
\centerline{\includegraphics[width=0.7\linewidth]{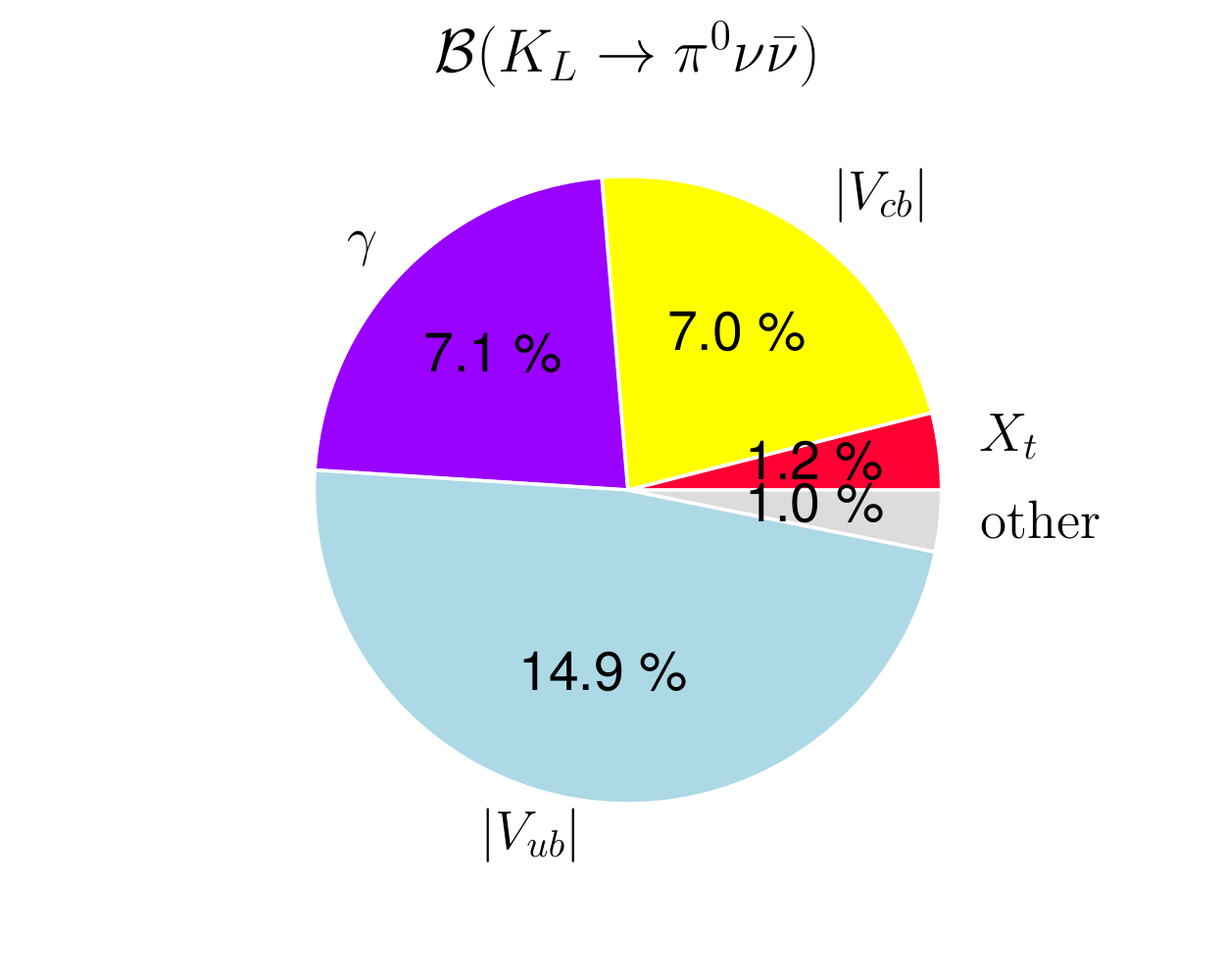}}
\end{minipage}
\caption[]{Error budgets for the branching ratio observables $\mathcal{B}(\kpn)$ and $\mathcal{B}(\klpn)$~\cite{Buras:2015qea}. 
The remaining parameters, which each contribute an error of less than 1\%, are grouped into the ``other'' category.}
\label{fig:pieplots}
\end{figure}

In the SM these decays are dominantly driven by so-called $Z$-penguins.
Due to the breaking of the GIM-mechanism by the squared masses of the internal quarks, the heavy top-quark loops dominate, yet charm-loops also remain relevant due to their larger CKM contribution.
The branching ratio observables can be expressed as
\begin{eqnarray}
    {\rm BR}(K^+ \to \pi^+ \nu\bar\nu) &=& \tilde\kappa_{+}\left[ 
        \left({{\frac{{\rm Im}(V_{td}V_{ts}^*)}{\lambda^5} X(x_t)}}\right)^2
    + \left({{\frac{{\rm Re}(V_{cd}V_{cs}^*)}{\lambda}P_c(X)}} + {{\frac{{\rm Re}(V_{td}V_{ts}^*)}{\lambda^5} X(x_t)}}\right)^2\right] \nonumber\\
    {\rm BR}(K_{\rm L} \to \pi^0 \nu\bar\nu) &=&
        \kappa_{\rm L} 
        \left({{\frac{{\rm Im}(V_{td}V_{ts}^*)}{\lambda^5} X(x_t)}}\right)^2
        \label{SMeqns}
\end{eqnarray}
where the accurately determined $\tilde{\kappa}_+$\footnote{The tilde denotes the inclusion of electromagnetic radiative correction.} and $\kappa_{\rm L}$ include the hadronic matrix elements~\cite{Mescia:2007kn}.
The charm loop contributions have been determined with NNLO QCD corrections~\cite{Buras:2005gr,Buras:2006gb} and NLO EW corrections~\cite{Brod:2008ss}. 
A numerical update gives~\cite{Buras:2015qea}
\begin{equation}
    {P_c(X)} = 0.404 \pm 0.024.
\end{equation}
Similarly, the top loops have been determined with NLO QCD corrections~\cite{Buchalla:1993bv,Misiak:1999yg} and NLO EW corrections~\cite{Brod:2010hi}, for which a numerical update gives~\cite{Buras:2015qea}
\begin{equation}
    {X(x_t)} = 1.481 \pm 0.005\big|_{\rm th} \pm 0.008\big|_{\rm exp}.
\end{equation}
That leaves only the CKM matrix element inputs $V_{td}V_{ts}^*$ and $V_{cd}V_{cs}^*$.

\begin{figure}[t]
\begin{minipage}{0.5\linewidth}
\centerline{\includegraphics[width=0.7\linewidth]{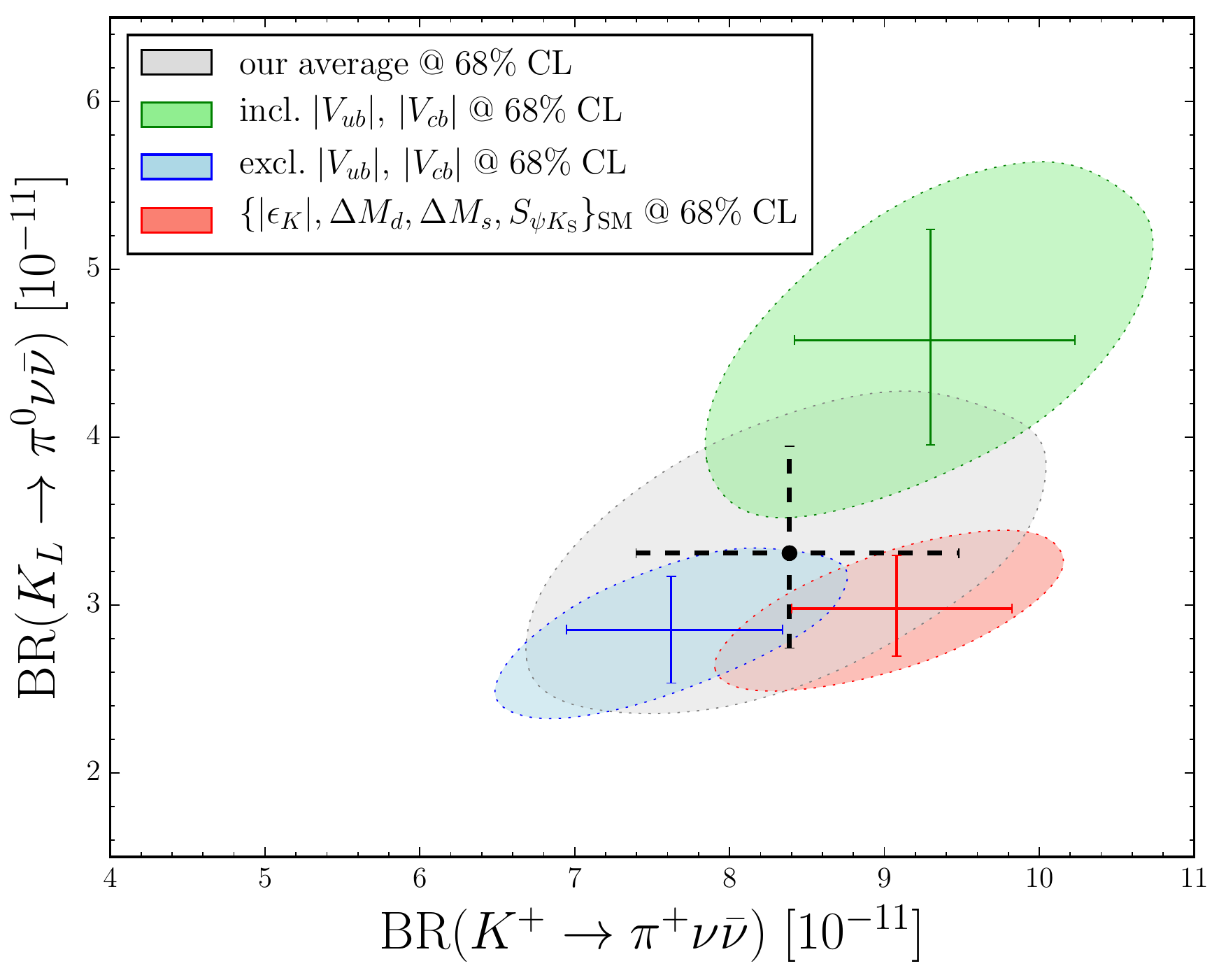}}
\end{minipage}
\hfill
\begin{minipage}{0.5\linewidth}
\centerline{\includegraphics[width=0.7\linewidth]{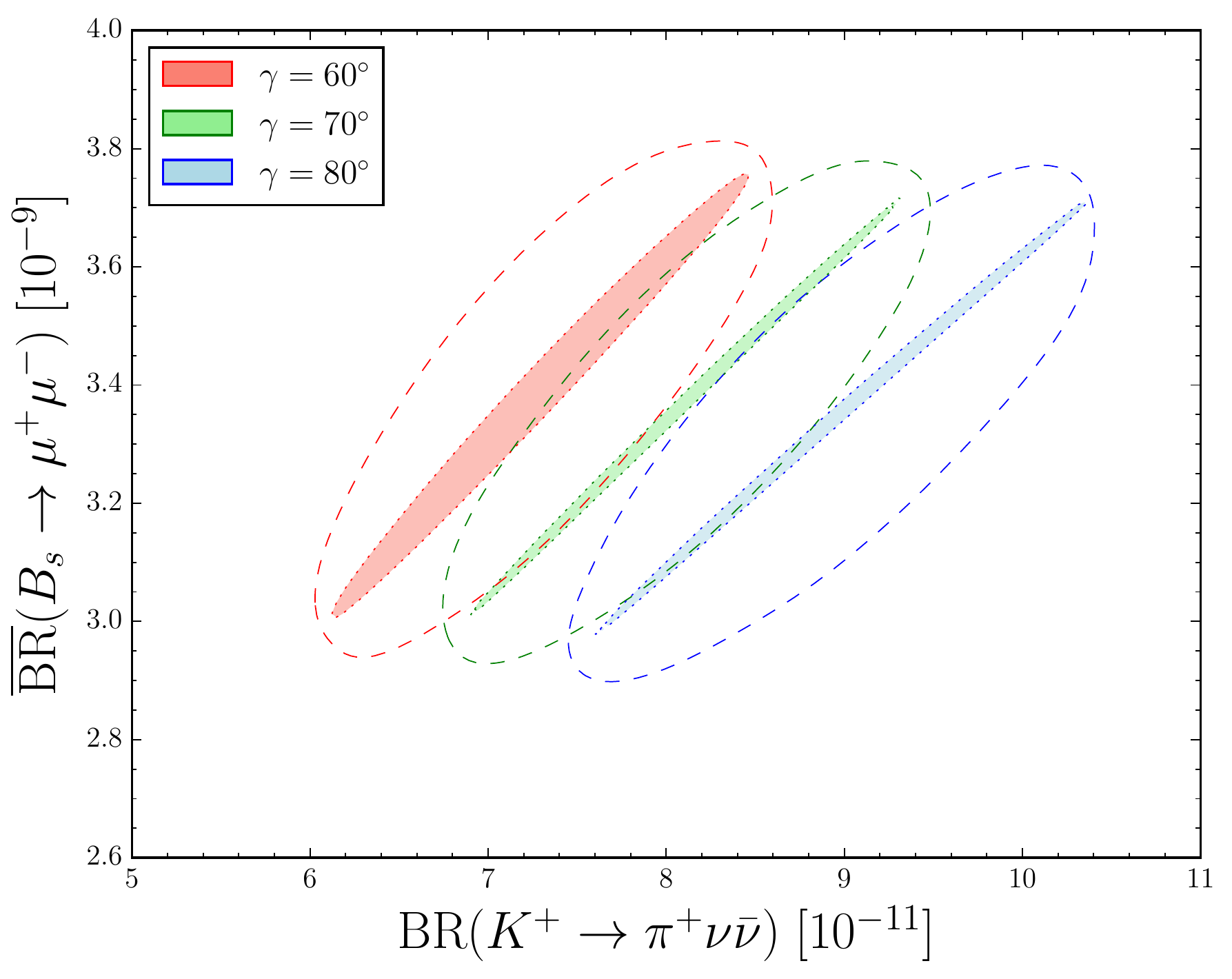}}
\end{minipage}
\caption[]{{\it Left panel:} comparison of 68\% C.L. regions for $\mathcal{B}(\klpn)$ and $\overline{\mathcal{B}}(B_s\to\mu^+\mu^-)$ versus $\mathcal{B}(\kpn)$, using different CKM inputs as described in the test~\cite{Buras:2015qea}. {\it Right panel:} correlation of $\overline{\mathcal{B}}(B_s\to\mu^+\mu^-)$ versus $\mathcal{B}(\kpn)$ for fixed values of $\gamma$: the dashed regions correspond to the 68\% C.L. from the uncertainties on all other inputs, 
while the inner filled regions correspond only to the uncertainties from the remaining CKM inputs~\cite{Buras:2015qea}.}
\label{fig:grandplots}
\end{figure}

For studies of NP it is preferable to use CKM inputs that are derived from tree-level observables,
namely from $|V_{us}|$, $|V_{ub}|$, $|V_{cb}|$ and the unitarity triangle angle $\gamma$, as these are not likely to be tainted by NP.
However, in doing so we encounter the currently large discrepancies between exclusive and inclusive determinations of the CKM matrix elements $|V_{cb}|$ and $|V_{ub}|$ from semileptonic $B$ decays (see~\cite{Ricciardi:2014aya}).
Specifically, we have $|V_{ub}|_{\rm excl} = (3.72 \pm 0.14)\times 10^{-3}$~\cite{Bailey:2014bea} versus $|V_{ub}|_{\rm incl} = (4.40 \pm 0.25)\times 10^{-3}$~\cite{Barberio:2007cr} and $|V_{cb}|_{\rm excl} = (39.36 \pm 0.75)\times 10^{-3}$~\cite{Aoki:2013ldr} versus $|V_{cb}|_{\rm incl} = (42.21 \pm 0.78)\times 10^{-3}$~\cite{Alberti:2014yda}.
This effect is unlikely to be due to NP~\cite{Crivellin:2014zpa}.
One way to proceed is to assume both determinations are equally correct and take a weighted average, inflating the errors via the PDG method~\cite{Beringer:1900zz}, which gives
\begin{equation}
    {|V_{ub}|_{\rm avg} = (3.88 \pm 0.29)\times 10^{-3},\quad 
    |V_{cb}|_{\rm avg} = (40.7 \pm 1.4)\times 10^{-3}}.
\end{equation}
Using these values together with $|V_{us}| = 0.2252\pm 0.0009$ and $\gamma = (73.2^{+6.3}_{-7.0})^\circ$~\cite{Trabelsi:2014} gives the branching ratio predictions
\begin{equation}
{\rm BR}(K^+\to \pi^+\nu\bar\nu)= (8.4 \pm 1.0)\times 10^{-11},\qquad
{\rm BR}(K_{\rm L}\to \pi^0\nu\bar\nu) = (3.4 \pm 0.6)\times 10^{-11}.
\end{equation}
In Figure~\ref{fig:pieplots} we show the corresponding error budgets, where the CKM errors are clearly seen to dominate both predictions.
The parametric dependence on the leading CKM input of both decays is given by~\cite{Buras:2015qea}
\begin{eqnarray}
    {\rm BR}(K^+\to \pi^+\nu\bar\nu) &=& (8.39\pm 0.30)\times 10^{-11} \left[\frac{|V_{cb}|}{40.7\times 10^{-3}}\right]^{2.8} \left[\frac{\gamma}{73.2^\circ}\right]^{0.708}\label{KpParam} \\
    {\rm BR}(K_{\rm L}\to \pi^0\nu\bar\nu) &=& (3.36\pm 0.05)\times 10^{-11} \left[\frac{|V_{ub}|}{3.88\times 10^{-3}}\right]^{2} \left[\frac{|V_{cb}|}{40.7\times 10^{-3}}\right]^{2} \left[\frac{\sin\gamma}{\sin(73.2^\circ)}\right]^{2} 
\end{eqnarray}
A comparison can be made with the CKM inputs determined purely from the loop-level observables $|\epsilon_K|$, $\Delta M_d$, $\Delta M_s$ and $S_{J/\psi K_{\rm S}}$, assuming no NP enters these observables~\cite{Buras:2015qea}.
The dominant uncertainty in this case is QCD lattice input, which, using the latest FLAG results~\cite{Aoki:2013ldr}, results in the more precise predictions
\begin{equation}
{\rm BR}(K^+\to \pi^+\nu\bar\nu)= (9.1 \pm 0.7)\times 10^{-11},\qquad
{\rm BR}(K_{\rm L}\to \pi^0\nu\bar\nu) = (3.0 \pm 0.3)\times 10^{-11}
\end{equation}
that are valid only in the SM.
In the left panel of Figure~\ref{fig:grandplots} a comparison of these results is given with the tree-level averages given above, as well as taking purely inclusive or exclusive values.

It is tempting to construct SM predictions independent of the tree-level $|V_{cb}|$ and $|V_{ub}|$ determinations.
To that end we can use that the $B_s\to\mu^+\mu^-$ branching ratio is effectively proportional to $|V_{cb}|^2$ -- its dominant uncertainty, followed by the $B_s$ meson decay constant $f_{B_s}$.
Combining this observable with (\ref{KpParam}) to eliminate $|V_{cb}|$, we then have to a very good accuracy in the SM the prediction~\cite{Buras:2015qea}
\begin{equation}
    {\rm BR}({K^+\to \pi^+\nu\bar\nu}) = (65.3\pm 2.9)\,\Big[\overline{\rm BR}({B_s\to\mu^+\mu^-}) \Big]^{1.4}\left[\frac{{\gamma}}{73.2^\circ}\right]^{0.708} \left[\frac{f_{B_s}}{227\,{\rm MeV}}\right]^{-2.8}
\end{equation}
We show this relation in the right panel of Figure~\ref{fig:grandplots} for fixed values of $\gamma$, and illustrate the small dependence on the remaining CKM inputs.

\section{$K\to \pi\nu\bar\nu$ beyond the Standard Model}\label{sec:BSM}

\begin{figure}
\begin{minipage}{0.5\linewidth}
\centerline{\includegraphics[width=0.7\linewidth]{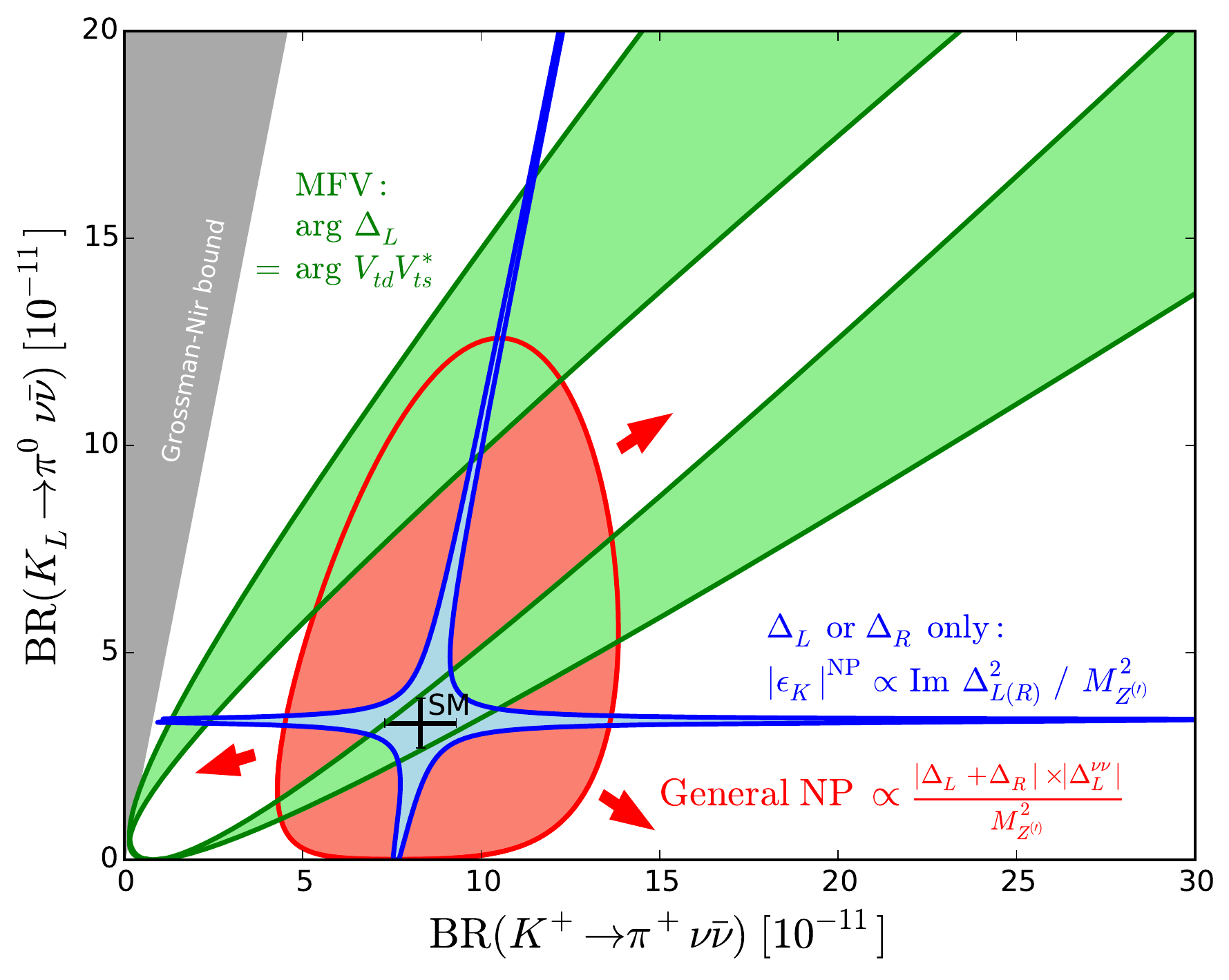}}
\end{minipage}
\hfill
\begin{minipage}{0.5\linewidth}
\centerline{\includegraphics[width=0.7\linewidth]{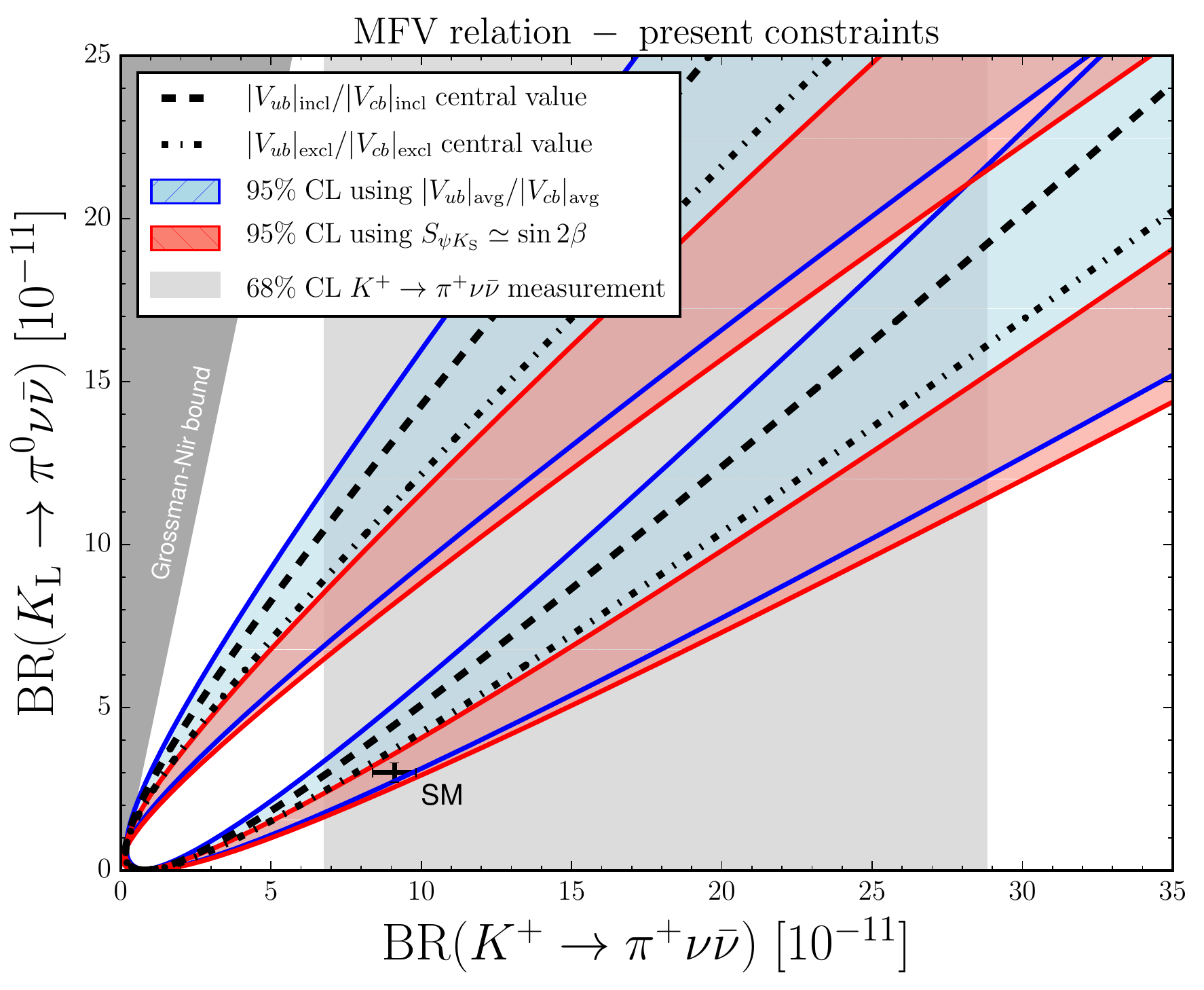}}
\end{minipage}
\caption[]{{\it Left panel:} illustration of coverage of general LH and RH NP with arbitrary CP phase (red region), compared with correlations occurring in the case of MFV (green region) or purely LH or RH NP subject to kaon mixing constraint (blue region). {\it Right panel:} the MFV relation between $\kpn$ and $\klpn$ using $S_{\psi K_{\rm S}}\simeq \sin 2\beta$ versus using the various tree-level inputs of $|{V_{cb}}/{V_{ub}}|$ and $\gamma$ compatible with current constraints.}
\label{fig:constPlots}
\end{figure}

Due to vanishingly small neutrino masses, Higgs-like scalar couplings to a pair of neutrinos are negligible both in and beyond the SM.
As a result NP contributions to $s\to d\nu\bar\nu$ transitions in the $K\to \pi\nu\bar\nu$ decays are typically mediated by vector bosons.
In this case NP generally enters in two ways: via modified $Z$ couplings to quarks, for example in the MSSM involving supersymmetric penguin processes, or via a new heavy $Z'$-like gauge boson.
To illustrate the main features of such models, we will consider simplified $Z$ and $Z'$ models with tree-level FCNC couplings to quarks -- for which we denote left and right handed couplings by $\Delta_{L,R}^{sd}(Z^{(\prime)})$ -- and diagonal coupling to neutrinos, denoted by $\Delta_L^{\nu\nu}(Z^{(\prime)})$.
The top-quark loop function in the SM then receives the following NP correction~\cite{Buras:2012jb}:
\begin{equation}
    {{X(x_t)}} \to X(x_t)_{\rm SM} + {\frac{\pi^2}{2 M_W^2 G_{\rm F}^2}}\frac{\Delta_L^{\nu\nu}(Z^{(\prime)})}{V^*_{ts}V_{td} M_Z^{(\prime)2}}\left[{{\Delta_L^{sd}(Z^{(\prime)})}} + {{\Delta_R^{sd}(Z^{(\prime)})}}\right],
\end{equation}
where $M_{Z'}$ is the mass of the heavy new $Z'$ boson.
From inspection of (\ref{SMeqns}) we observe that $K^+\to \pi^+\nu\bar\nu$ is sensitive to both the real and imaginary NP contributions to ${X(x_t)}$, while $K_{\rm L}\to \pi^0\nu\bar\nu$ only to the latter.
In the left panel of Figure~\ref{fig:constPlots} the red region illustrates the general coverage of left and right handed NP with an arbitrary CP violating phases in the $K^+\to \pi^+\nu\bar\nu$ versus $K_{\rm L}\to \pi^0\nu\bar\nu$ plane i.e.\ in general there is no correlation present.

Minimal Flavour Violation (MFV) is a mechanism to protect against large FCNCs in models beyond the SM by insisting that FCNCs can only arise from SM Yukawas.
For the decays in question this implies the combination $V_{td}V_{ts}^*\,X(x_t)$ can be modified by NP provided any shifts with respect to $X(x_t)$ are real valued.
Translated to our simplified models this means ${\rm arg}(\Delta_L^{sd}) = {\rm arg}(V_{td}V_{ts}^*)$ and that $\Delta_R^{sd} = 0$ for both $Z$ and $Z'$.
The condition of MFV results in the green band shown in the left panel of Figure~\ref{fig:constPlots}.
The CKM input entering this correlation is to a very good accuracy only the UT angle $\beta$, which gives a triple correlation between these two decays and the CP violating observable $S_{J/\psi K_{\rm S}}$ that is unaffected by NP in MFV~\cite{Buras:2001af}.
In the right panel of Figure~\ref{fig:constPlots} we illustrate the current status of this relation, and compare it to the larger errors obtained from using tree-level CKM inputs.

Correlations between the $K\to \pi\nu\bar\nu$ decays also arise from considering the constraints from other kaon observables.
Notably NP contributions to CP violation in kaon mixing are proportional to~\cite{Buras:2014zga}
\begin{equation}
    \epsilon_K^{\rm NP} \propto \frac{1}{M_Z^{(\prime)2}}\,{\rm Im}\left[ \Delta_L^{sd}(Z^{(\prime)})^2
    + \Delta_R^{sd}(Z^{(\prime)})^2 
    + 2 \kappa_{sd}\,
    \Delta_L^{sd}(Z^{(\prime)})
    \Delta_R^{sd}(Z^{(\prime)}) \right],
\end{equation}
where $\kappa_{sd}$ is the ratio of the corresponding hadronic matrix elements.
Thus in the case of purely left or right handed NP, the square of the respective imaginary components are strongly constrained by experiment, which results in the two blue branches illustrated in the left panel of Figure~\ref{fig:constPlots}.
Aside from its presence in various $Z'$ models with purely left or right handed couplings~\cite{Buras:2012dp,Buras:2012jb}, this correlation also appears for instance in Little Higgs models with T-parity~\cite{Blanke:2009am}.
In a Randall-Sundrum model with a custodial symmetry that allows only for large new right-handed FCNC couplings, the correlation is lost due to the kaon mixing constraint being saturated by additional NP~\cite{Blanke:2008yr}.

\begin{figure}
\begin{minipage}{0.5\linewidth}
\centerline{\includegraphics[width=0.7\linewidth]{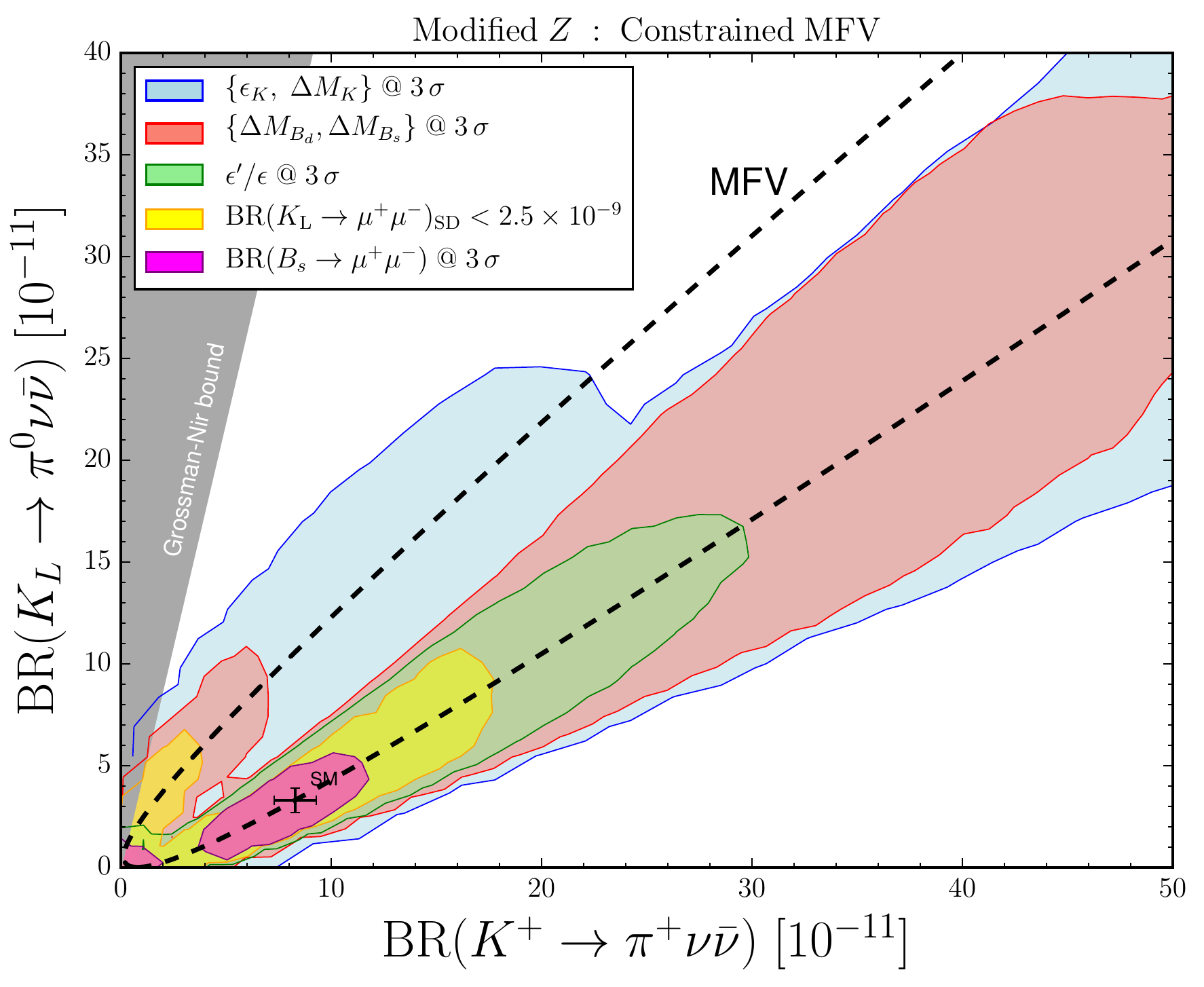}}
\end{minipage}
\hfill
\begin{minipage}{0.5\linewidth}
\centerline{\includegraphics[width=0.7\linewidth]{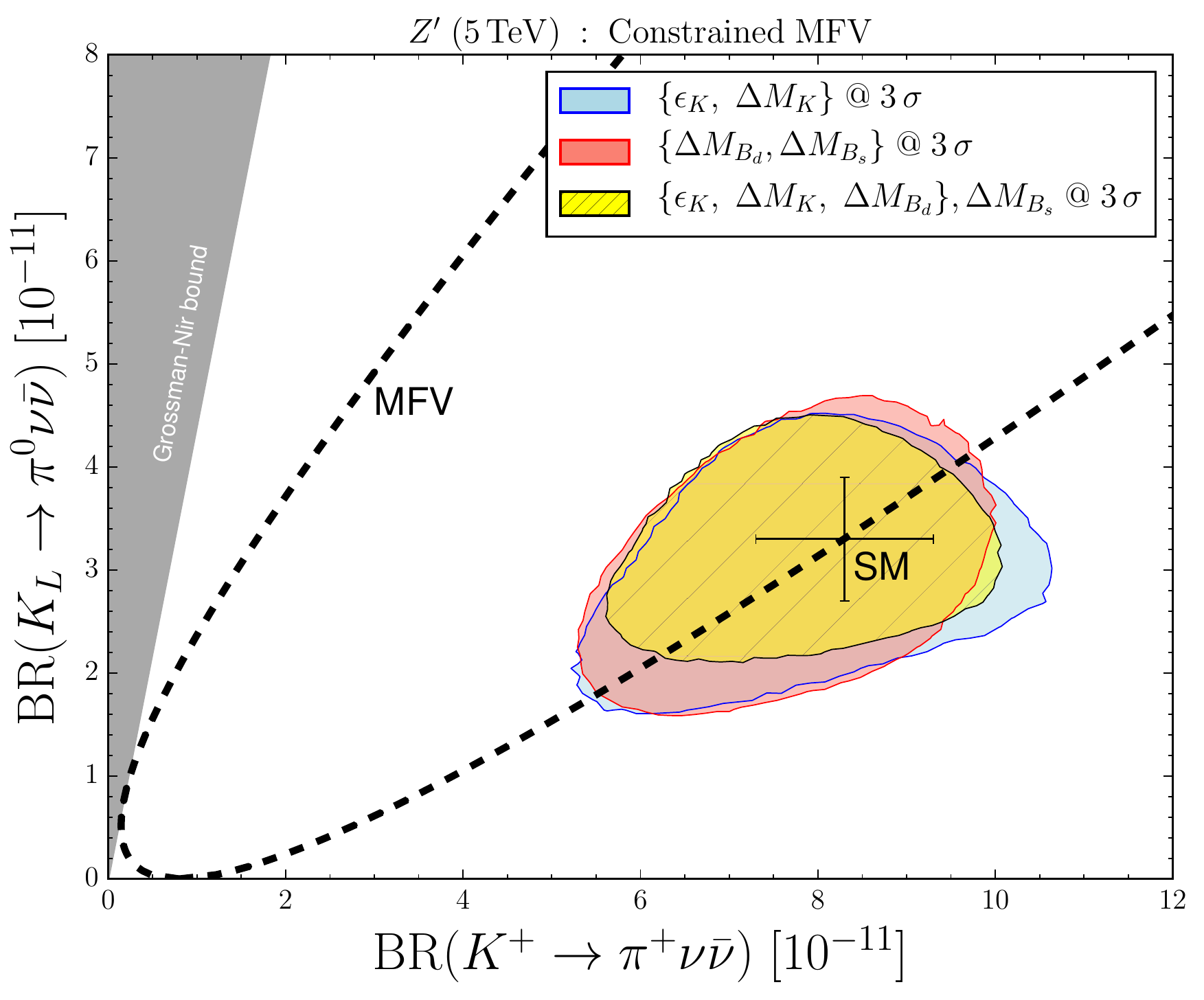}}
\end{minipage}
\caption[]{The $3\,\sigma$ allowed ranges for  $\mathcal{B}(\klpn)$ and $\mathcal{B}(\kpn)$ in a simplified model with modified $Z$ couplings (left panel) or a simplified model with a $5\tev~Z'$ (right panel) obeying CMFV~\cite{Buras:2015inprep}. In the former case $\Delta F =1$ constraints are most constraining, while in the latter $\Delta F =2$ are very constraining assuming the $Z'$ has the same coupling to neutrinos as the $Z$.}
\label{fig:MFV}
\end{figure}

Besides from kaon mixing, also direct CP violation in $K\to\pi\pi$, namely $\epsilon'/\epsilon$, can lead to strong constraints for the imaginary parts of $\Delta_{L,R}^{sd}$.
In this case the imaginary components are not squared, so that limits on $\epsilon'/\epsilon$ directly limit the branching ratio of $K_{\rm L}\to \pi^0\nu\bar\nu$.
This is in particular the case for a simplified $Z$ model, as in a simplified $Z'$ model the diagonal quark couplings must also be addressed for this constraint to be meaningful.
If the coupling of the vector boson driving the $s\to d\nu\bar\nu$ transition is related to $s\to d\mu\bar\mu$, for example by $SU(2)_L$ symmetry, then also the current upper bound ${\rm BR}(K_{\rm L} \to \mu^+\mu^-)_{\rm SD} < 2.5\times 10^{-9}$~\cite{Isidori:2003ts} can constrain NP in ${\rm BR}(K^+\to \pi^+ \nu\bar\nu)$, as its NP contribution is proportional to ${\rm Re}(\Delta_L^{sd} - \Delta_R^{sd})$.
The (anti-)correlation of these two branching ratios can also reveal the presence of left-handed (right-handed) NP~\cite{Buras:2013ooa}, as for example demonstrated in Randall-Sundrum~\cite{Blanke:2008yr} and partially composite models~\cite{Straub:2013zca}.

In models with MFV the corresponding CKM suppression usually implies that the NP effects compatible with present constraints are small.
In the MSSM with MFV, for example, it has been shown that NP effects are in general limited to be about 10\%~\cite{Isidori:2006qy,Smith:2014mla}.
In the left panel of Figure~\ref{fig:MFV} we show the applicable constraints in the $K^+\to \pi^+\nu\bar\nu$ versus $K_{\rm L}\to \pi^0\nu\bar\nu$ plane for a simplified $Z$ model with MFV~\cite{Buras:2015inprep}.
Similarly, in the right panel we show the constraints for a simplified $Z'$ obeying MFV, with a mass of $5\,{\rm TeV}$ and the same strength couplings to neutrinos as the $Z$~\cite{Buras:2015inprep}.
We observe that in the case of the lighter $Z$ the $\Delta F=1$ constraints are the most constraining, while for the $Z'$ the $\Delta F=2$ constraints are already very constraining.

Finally let us address what NP scales could ultimately be reached, taking as a benchmark a simplified $Z'$ model with maximum couplings to quarks and leptons consistent with perturbativity.
If the $Z'$ couples only left or right handedly to quarks, the constraints from kaon mixing apply, and scales as high as 50~TeV, or equivalently 4 zeptometers can be reached~\cite{Buras:2014zga}.
If both LH and RH couplings are present, then a tuning is possible that cancels the kaon mixing constraint, allowing distances under a zeptometer to be probed.
In other words, the $K\to\pi\nu\bar\nu$ decays will not only be excellent probes of NP in the coming decade, they could eventually allow the {\it Zeptouniverse} to be probed by $\Delta F=1$ rare decays processes.


\section*{Acknowledgments}

I would like to thank the organizers of the Moriond Electroweak Conference for their organization and hospitality.
I am grateful to Andrzej Buras, Dario Buttazzo and Jennifer Girrbach-Noe for their collaboration on topics presented in this talk.
This research was completed and financed in the context of the ERC Advanced Grant project ``FLAVOUR''(267104) and was partially supported by the DFG cluster of excellence ``Origin and Structure of the Universe''.

\section*{References}

\bibliography{allrefs}

\end{document}